# Light-matter interaction in a microcavity-controlled graphene transistor

Michael Engel[1,2], Mathias Steiner[3,*], Antonio Lombardo[4], Andrea C. Ferrari[4], Hilbert v. Löhneysen[2,5,6], Phaedon Avouris[3], and Ralph Krupke[1,2,7,#]

[1]*Institute of Nanotechnology, Karlsruhe Institute of Technology, 76021 Karlsruhe, Germany*

[2] *DFG Center for Functional Nanostructures (CFN), 76031 Karlsruhe, Germany*

[3] *IBM Thomas J. Watson Research Center, Yorktown Heights, New York 10598, USA*

[4] *Engineering Department, University of Cambridge, Cambridge CB3 0FA, UK*

[5] *Physikalisches Institut, Karlsruhe Institute of Technology, 76031 Karlsruhe, Germany*

[6] *Institute of Solid State Physics, Karlsruhe Institute of Technology, 76021 Karlsruhe, Germany*

[7] *Institut für Materialwissenschaft, Technische Universität Darmstadt, 64287 Darmstadt, Germany*

*msteine@us.ibm.com, #krupke@kit.edu

**Abstract:**

Graphene has extraordinary electronic and optical properties and holds great promise for applications in photonics and optoelectronics. Demonstrations including high-speed photodetectors, optical modulators, plasmonic devices, and ultrafast lasers have now been reported. More advanced device concepts would involve photonic elements such as cavities to control light–matter interaction in graphene. Here we report the first monolithic integration of a graphene transistor and a planar, optical microcavity. We find that the microcavity-induced optical confinement controls the efficiency and spectral selection of photocurrent generation in the integrated graphene device. A twenty-fold enhancement of photocurrent is demonstrated. The optical cavity also determines the spectral properties of the electrically excited thermal radiation of graphene. Most interestingly, we find that the cavity confinement modifies the electrical transport characteristics of the integrated graphene transistor. Our experimental approach opens up a route towards cavity-quantum electrodynamics on the nanometre scale with graphene as a current-carrying intra-cavity medium of atomic thickness.

Graphene, an atomic monolayer formed by carbon hexagons, is a material with extraordinary electrical and optical[1–3] properties. Consequently, there is a growing interest in graphene optoelectronics[4] and first demonstrations of graphene-based photodetectors[5], optical modulators[6], plasmonic devices[7,8] and ultra-fast lasers[9] have been reported. Because of its two- dimensional geometry, graphene is ideally suited for enclosure within a planar λ/2 microcavity, a photonic structure that confines optical fields between two highly reflecting mirrors with a spacing of only one half wavelength of light. The optical confinement could provide a powerful means of controlling the otherwise featureless optical absorption[10] as well as the spectrally broad thermal emission[11,12] of graphene. The concept of optical confinement of graphene enables a new class of functional devices as, for example, spectrally selective and highly directional light emitters, detectors and modulators. Moreover, it opens up the opportunity for investigating fundamental, cavity-induced modifications of light–matter interaction in graphene.

According to Fermi's golden rule , the spontaneous photon emission (absorption) rate is determined by the local photonic mode density that can be significantly altered inside an optical microcavity[14]. The cavity-induced confinement enables the enhancement[15] or the inhibition[16] of the light emission (absorption) rate of the intra-cavity medium. The in-plane transition dipole moment of the intra-cavity medium couples to the longitudinal cavity mode of the cavity with wavelength-dependent efficiency, and the coupling strength is maximized at the antinode of the optical field located at the cavity centre[17,18]. Figure 1 visualizes the principle of confining graphene by a planar optical cavity.

The cavity-induced rate enhancement, or Purcell-effect, has already been demonstrated with quasi 2D quantum wells, as well as quasi 1D and 0D systems such as atoms, molecules, quantum dots and other nanoparticles[14]. However, embedding a truly 2D material such as graphene into a planar cavity has not been reported so far. Yet such an approach is highly desirable for two main reasons. First, the coupling area, that is, the spatial overlap between graphene and the cavity, can be extended to the micrometre scale within the two dimensions of the cavity plane, whereas preserving the optical confinement of graphene with respect to the cavity normal on the length scale of $\lambda/2$ (see Fig. 1). This is important because optical transitions in graphene are associated with in-plane transition dipole moments ($\pi$–$\pi$* transitions), thus rendering the 1D planar cavity confinement highly efficient. Second, the Fermi energy and density-of-states of graphene can be tuned easily by directly connecting it to electrodes, thus enabling charge carrier control, electrical transport and heating within the active area of the cavity. Graphene hence opens a unique experimental approach towards cavity-quantum electrodynamics on the nanometre scale with a current-carrying intra-cavity medium of atomic thickness.

In this paper, we report the monolithic integration of a graphene transistor with a planar optical microcavity. We find that both photocurrent generation as well as electrically excited thermal light emission of graphene can be controlled by the spectral properties of the microcavity. The device constitutes a first implementation of a cavity-enhanced graphene light detector, as well as a demonstration of a fully integrated, narrow-band thermal light source. Most importantly, the optical confinement of graphene by the microcavity profoundly modifies the electrical transport characteristics of the integrated graphene transistor.

**Results**

**Device design.** Figure 2a outlines our novel integration principle: an electrically contacted single layer of graphene is embedded between two optically transparent, dielectric thin films made of $Si_3N4$ and $Al_2O_3$, respectively. The dielectric layers are enclosed by two metallic (Ag, Au) mirrors with a spacing L that determines the resonance wavelength $\lambda_{cavity}$ of the microcavity (see Methods and ref. 18). A novel multi-step manufacturing process (see Methods) allows us to define device area and cavity mirror spacing with nanometre precision and to build a series of devices that satisfy the specific requirements of different optoelectronic experiments. For photocurrent studies, we designed the cavity resonance to match the tuning range of the laser system at hand (~580nm). For thermal emission studies, we designed the cavity resonance to be spectrally located within the detection range of the spectroscopic unit (~925nm). In principle, all the measurements can be performed on the same device, whereas the efficiencies of light absorption and emission are largely determined by the optical properties of the cavity.

**Microcavity-controlled light detection**. Figure 2b–e reports the device characterization by electrical, optical and optoelectronic measurements: to characterize the electronic device properties, we measure both electrical transfer and output characteristics of the integrated graphene transistor (Fig. 2e). We apply a bias voltage along the graphene sheet and use one of the metallic cavity mirrors as a gate electrode (Fig. 2a). We fit the measured electrical transfer characteristics based on the model reported in ref. 19 and extract the device parameters that demonstrate the quality of the graphene sheet and the Pd-graphene contacts: we find a carrier mobility $\mu=2350 \text{cm}^2\text{V}^{-1}\text{s}^{-1}$, a residual carrier density $n_0=4.9\times10^{11}\text{cm}^{-2}$, an electrical-resistance ratio resistance $R_{max}/R_{min}=5$, and a specific contact $R_c=0.265\text{k}\Omega\mu\text{m}$.

To characterize the optical properties of the device, we illuminate the cavity with white light from the top (see Fig. 2a) and spectrally analyse the transmitted light. This way, we determine the resonance wavelength $\lambda_{cavity}$ and the cavity quality factor $Q=\lambda_{cavity}/\Delta\lambda_{cavity}$, where $\Delta\lambda_{cavity}$ is the spectral full-width-at-half-maximum (FWHM) of the peak at λcavity in the measured transmission spectrum. In the present case (Fig. 2d), we obtain $\lambda_{cavity}=585\text{nm}$ and cavity-Q=20.

We employ photocurrent generation in graphene[20] to probe electronically the optical absorption of the graphene layer inside the device. We focus a laser beam on one of the microcavity mirrors, tune the laser wavelength across the optical resonance, and measure the photocurrent generated inside the biased graphene layer by means of a lock-in technique. The functional dependence of the photocurrent amplitude on laser wavelength (Fig. 2d, red circles) matches the spectral profile of the cavity resonance as measured by white light transmission micro-spectroscopy (Fig. 2d, solid line). By tuning the laser wavelength to $\lambda_{laser}=583\text{nm}$, on resonance with the microcavity, we obtain a photocurrent amplitude of 23.3nA, while we measure only 1.2nA for laser illumination at $\lambda_{laser}=633\text{nm}$. The photocurrent modulation by a factor of 20 demonstrates that the device acts as a light detector with spectral selectivity. In contrast, the photocurrent amplitude of a non-confined graphene tran- sistor varies by less than a factor of two if we tune the laser excitation wavelength across the same spectral interval.

**Microcavity-controlled thermal light emission.** We now investigate the light-emission properties of non-confined graphene and compare it to the microcavity-controlled graphene transistor. By applying a bias voltage across the non-confined graphene device using source and drain electrodes (Fig. 3a), the electrical current heats up the graphene layer and thermal light emission sets in, as reported previously[11,12]. As shown in Fig. 3b, the electrically excited, thermal emission spectrum of graphene in free, non-confined space exhibits a featureless exponential tail that shifts from the near-infrared towards the visible spectral range as a function of injected electrical power. The temperature values are extracted by fitting the measured emission spectra based on a model of a two-dimensional black body (see Methods and ref. 21).

In contrast, the thermal emission spectrum of a microcavity-controlled graphene transistor displays a single, narrow peak at $\lambda_{cavity}$=925nm having a FWHM of 50nm (Fig. 3b), providing a 140-fold spectral narrowing as compared with the simulated free-space thermal spectrum at $T$=650K. It is important to note that this is not merely a spectral filtering effect, but that thermal radiation cannot be emitted by the graphene layer at all if the thermal radiation wavelength $\lambda_{thermal}$ is larger than $\lambda_{cavity}$ because of the optical confinement, or, in other words, the cavity-induced inhibition of spontaneous emission. This constitutes the first demonstration of a current-driven, microcavity-controlled thermal light source. The spectral peak position of the emission peak does not shift as a function of injected electrical power. The simulated emission spectra in Fig. 3c reproduce the overall shape of their experimental counterparts very well (for details, see Methods). Further data analysis reveals that microcavity-controlled thermal radiation is emitted into a narrow lobe with a total angular width of 24° (FWHM) only (see Supplementary Methods and Supplementary Fig. S1). The integral over the microcavity-controlled light intensity is plotted as function of the injected electrical power in Fig. 3d for three different devices. The power dependences reveal that the integrated light intensity is proportional to $T^3$ as expected from the Stefan-Boltzmann law in two dimensions[21]. Here we assume that the electrical power density $p \propto T$, which is validated by measurements in non-confined space (see Supplementary Fig. S2).

We now establish the physical concept of cavity-controlled thermal light generation in graphene. The graphene layer initially heats up due to carrier scattering. The emission of long-wavelength thermal radiation at $\lambda_{thermal} > \lambda_{cavity}$ is however inhibited by the cavity. As the temperature in the graphene sheet increases as a function of the electrical power density $p$, the thermal distribution inside the graphene layer spectrally shifts towards $\lambda_{cavity}$ (Fig. 3b) and, at a threshold temperature $T_{cavity}$, eventually enables the emission of a significant fraction of photons at $\lambda_{thermal} \lesssim \lambda_{cavity}$. We note that the temperature $T_{cavity}$ and the corresponding threshold power density $p_{cavity}$ depend on device parameters, that is, the carrier mobility $\mu$, the channel area and $\lambda_{cavity}$. For the device in Fig. 3b, we find $T_{cavity} \approx 650K$ and $p_{cavity} = 90 kWcm^{-2}$. Based on Wien's law in two dimensions,[21] we estimate that the graphene layer would have to be heated up to $T_{cavity,optimum} \approx hc/3.92 k_B \lambda_{cavity} \approx 4,000K$ to maximize the light output at $\lambda_{cavity} = 925nm$. In this case, the intensity maximum of the thermal-radiation distribution would overlap with the cavity resonance at $\lambda_{cavity}$, resulting in a peak population of the cavity mode by thermal photons. Note that we designed microcavity devices specifically for performing the thermal emission spectroscopy in the near-infrared to take advantage of the favourable experimental conditions, that is, the high sensitivity and detection yield of the CCD array at hand. The optimum light-emitting performance, however, is expected to be in the mid-infrared spectral range. Assuming device operation at, for example, $T_{cavity} = 650K$, we estimate a maximum thermal light output at $\lambda_{thermal} \approx 5.6\mu m$. Future work should hence extend the spectral range towards the terahertz regime.

**Transport in microcavity-controlled graphene transistors.** Finally, we discuss modifications of electrical transport in optically confined graphene by correlating the electrical current and the spectrally integrated thermal radiation.

In Fig. 4a–c, we show the electrical output characteristics of a graphene transistor in non-confined space. The electrical current in the graphene layer saturates in the high bias regime, whereas thermal light emission sets in as it shifts into the detection range. Performing the same experiment for a cavity-controlled graphene transistor with identical dimensions, we find that the electrical output characteristic is qualitatively different (see Fig. 4d–f) and we can identify three different regimes by comparing electrical transport and light emission properties. In the sub-threshold regime (I), the electrical current saturates and the graphene layer heats up while the cavity-induced inhibition of spontaneous emission for $\lambda_{thermal} > \lambda_{cavity}$ prevents off-resonant thermal radiation. In the threshold regime (II), the temperature of the graphene sheet has reached the critical value Tcavity enabling light emission at $\lambda_{thermal} \lesssim \lambda_{cavity}$. In the above-threshold regime (III), the initial electrical current saturation is lifted and the electrical resistance drops as function of electrical power density.

**Discussion**

We have observed electrical transport modifications similar to those shown in Fig. 4 in all functional graphene-cavity devices (a total of five), that is, those that allowed for thermal light generation (see Supplementary Fig. S3). In all cases, the saturation currents in regime I were lower than those obtained in the non-confined reference devices (same contacts and dielectric layers, no metal mirrors). One can rationalize this observation by assuming that the onset of current saturation in graphene depends on temperature and that the degree of self heating is determined by the thermal coupling of graphene to its local environment, captured by the thermal conductance $r$ as suggested in ref. 22. The saturation current $j$ in the graphene layer would then be proportional to $r$. This implies that cavity-induced variations of $r$ with respect to the non-confined reference value $r_0$ will lead to variations of the saturation current $j$ and, accordingly, to the temperature in the graphene layer. In this scenario, the cavity-induced inhibition of the radiative thermal relaxation leads to enhanced self heating of the graphene layer and an onset of current saturation at lower electrical power levels as compared with the non-confined case. Based on the experimental transport data and the self-heating model presented in ref. 22, we estimate temperature changes as high as $\Delta T=100K$ when compared with the same graphene transistor in non-confined space (see Supplementary Methods and Supplementary Fig. S4). In this context, we point out that the amount of heat that is radiatively dissipated from the graphene sheet is very small, orders of magnitude smaller than the amount of energy that is dissipated non-radiatively[11] (see Supplementary Fig. S5). It is clear then that a proper model of electrical transport and its temperature dependence in a microcavity-controlled graphene transistor should account for the current-induced self-heating of graphene that is affected by both the non-radiative heat transfer through dielectric interfaces and metal contacts, and the microcavity-controlled, radiative heat transfer.

In summary, we have demonstrated that a microcavity-controlled graphene transistor can act as a spectrally selective light detector and emitter with greatly enhanced sensitivity. Moreover, we have found that the cavity-induced optical confinement modifies graphene's electrical transport characteristics, an effect that may have important implications for nanoelectronics as well as cavity quantum electrodynamics.

**Methods**

**Device fabrication**

Graphene sheets are produced via micromechanical cleavage of graphite on Si substrates covered with 300nm of $SiO_2$ layer[23]. Single-layer graphene is identified by a combination of optical microscopy and Raman spectroscopy[24,25]. Three layers of 950K poly(methyl methacrylate) (PMMA) are then spin coated on the substrates where flakes are deposited. The samples are subsequently immersed in de-ionized (DI) water at 90°C for 2h, resulting in the detachment of the polymer film, due to the intercalation of water at the polymer–$SiO_2$ interface. Graphene flakes stick to the PMMA film, and can thus be removed from the original substrate. The target substrate is a suspended $Si_3N_4$ layer [$n(Si_3N_4)\approx2$] with a thickness of 50nm and an area of 50×50μm$^2$, supported by a Si frame having a thickness of 200μm. We defined metallic markers by e-beam lithography and e-beam evaporation of 5nm Ti and 50nm Au on the target substrate. These markers are used for orientation during the transfer process and re-alignment for all following e-beam lithography steps. The PMMA+graphene film is transferred onto the suspended $Si_3N_4$ layer. Because a thin layer of water is trapped at the substrate–polymer interface, the latter can be moved across the target substrate allowing accurate positioning of a chosen graphene flake onto a specific location on the $Si_3N_4$ membrane. The sample is then left to dry, and finally PMMA is dissolved by acetone drop casting followed by immersion, resulting in the gentle release of the selected graphene flake on the target substrate. Success of the transfer is confirmed by Raman spectroscopy, which also proves the absence of process-induced structural defects. Metallic contacts are fabricated by e-beam lithography and sequential deposition of 0.5nm Ti and 50nm Pd by e-beam evaporation. In a next step, the selected single-layer graphene are shaped by oxygen plasma etching into different sizes ($0.5\times0.5\mu m^2$, $1\times1\mu m^2$, $2\times2\mu m^2$ and $4\times4\mu m^2$). We then deposit a nucleation layer of 2nm Al on top of the single-layer graphene to ensure homogeneous growth[26], followed by an $Al_2O_3$ layer ($n(Al_2O_3)\approx1.7$) grown by atomic layer deposition of varying thickness. The chosen thicknesses of the intra-cavity dielectrics determine the resonance wavelength of the optical microcavity. To make the devices electrically accessible, we pattern openings over the large contact pads by e-beam lithography and etch away the $Al_2O_3$ in 40% phosphoric acid by weight at a temperature of $T$=50°C (etch rate ~10nm/2min) utilizing the PMMA resist as the etch mask. Cavity mirrors are prepared by depositing a 30nm Ag (Au) globally on the backside of the sample by e-beam evaporation. An additional e-beam lithography step is necessary for the local definition of the top cavity mirror followed by deposition of 60nm Ag (Au) to ensure devices can still be addressed electrically (see Supplementary Figs S6 and S7).

**Simulation of microcavity spectra**

Transmission and reflection spectra of the planar microcavity are simulated by using a transfer matrix method for multilayer stacks[27]. The stack used in the simulation consists of Ag(60nm)/$Al_2O_3$(0-100nm)/$Si_3N_4$(50nm)/Ag(30nm). The simulations reproduce well the experimental results obtained from a set of reference samples (see Supplementary Fig. S8).

The thermal emission spectra of graphene in free, non-confined space are fitted with the spectral density model of the two-dimensional black-body radiation given in ref. 21 and the temperature is extracted for each electrical power density level. The microcavity-controlled thermal emission spectra of graphene are then simulated for the same temperatures by superimposing a single Lorentzian at $\lambda_{cavity}$=925nm with a FWHM of 30nm on the spectral density of the two- dimensional black-body radiation discussed above. Note that, as compared with the measured white-light transmission spectrum of the same device, the peak in the thermal emission spectrum is blue shifted by 20nm and broadened by a factor of 2 (see Supplementary Fig. S1). This is mainly due to the detection with high numerical aperture, NA=0.8, leading to wavelength-dependent variations of the collection efficiencies for on- and off-axis cavity emissions[18], an effect that is not accounted for in the present model.

**Acknowledgment**

We acknowledge S. Linden (University of Bonn) for support with the optical simulations, D. B. Farmer (IBM TJ Watson Research Center) for atomic layer deposition, R. Ferlito (CSS, IBM TJ Watson Research Center) for Ag mirror deposition and B. A. Ek (IBM TJ Watson Research Center) for expert technical assistance. M.E. and R.K. acknowledge support by the Deutsche Forschungsgemeinschaft (DFG) and the State of Baden-Württemberg through the DFG-Center for Functional Nanostructures (CFN) within subproject B 1.9. A.C.F. acknowledges funding from EU grants NANOPOTS, GENIUS, RODIN and EPSRC grants EP/GO30480/1 and EP/G042357/1, Royal Society Wofson Research Merit Award.

**Author Contributions**

The experiments were conceived, designed, and carried out by M.E. and M.S.. A.L. performed graphene exfoliation and transfer. The manuscript was written by M.E. and M.S. with input from R.K., P.A., A.C.F., and H.v.L..

**Note added in proof:** A photodetection study of graphene inside a planar optical microcavity was reported while this work was under consideration (Furchi, M. et al. Microcavity-integrated graphene photodetector. Nano Lett. doi: 10.1021/nl204512x (2012)).

**Competing Financial Interests**

The authors declare no competing financial interests.

**Figures**

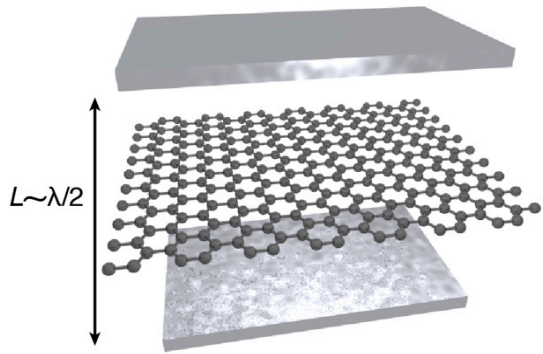

**Figure 1 | Microcavity-induced optical confinement of graphene.** Visualization of a graphene layer located at the center of a planar optical $\lambda/2$ microcavity. Optical fields with wavelength $\lambda$ are confined in the direction perpendicular to the cavity mirrors with spacing $L$. The optical coupling is maximized if the graphene layer is oriented parallel to the cavity mirrors.

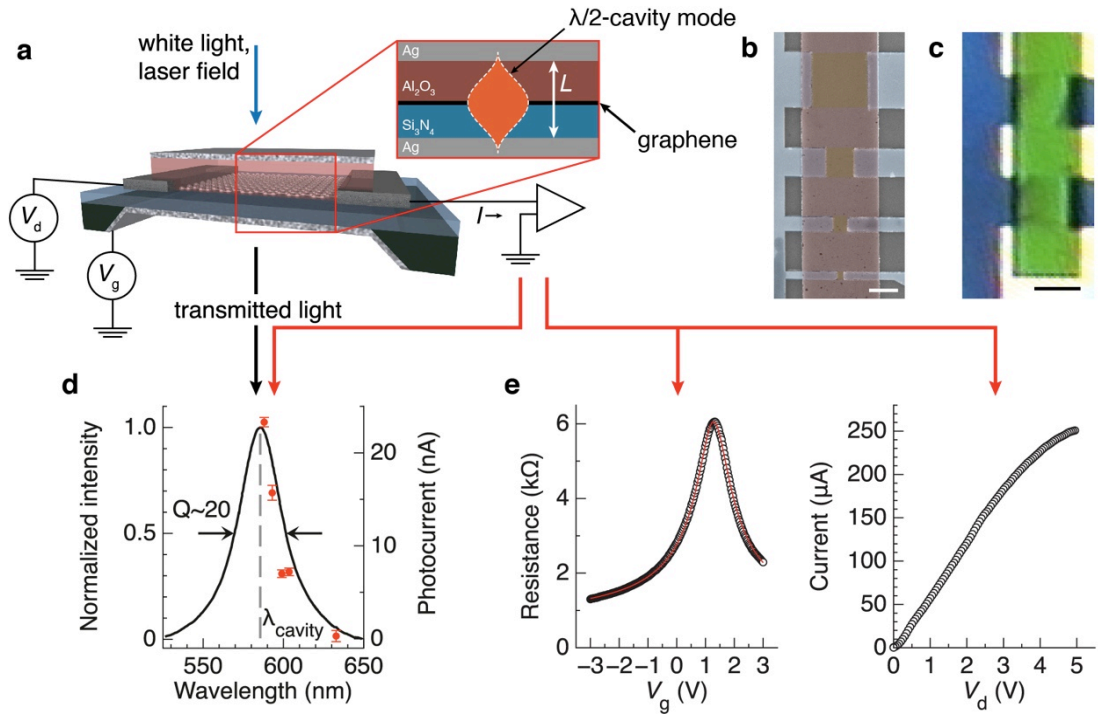

**Figure 2 | Microcavity-controlled graphene transistor and photocurrent generation. a**, Schematic representation and electrical interconnection of the device. Inset: Cross sectional view of the device. The graphene sheet is embedded between two Ag mirrors and separated by two dielectric layers ($Si_3N_4$; $Al_2O_3$). The thickness $L$ of the dielectric stack between the cavity mirrors determines the resonance wavelength λ of the optical microcavity. Also shown is a visualization of the intensity profile of the fundamental λ/2 cavity mode. **b**, Top-view scanning electron microscope false color image of the device; graphene sheets (yellow), Pd contacts (blue) Ag mirror (red). Scale bar, 2μm. **c**, Optical white light transmission micrograph of the device. The fundamental cavity mode is spectrally located at $\lambda_{cavity}$=585nm which appears green to the eye. Scale bar, 4 μm. **d**, Optical transmission spectrum of the device (black line) measured with white light illumination reveals the cavity resonance at $\lambda_{cavity}$=585nm having a cavity-$Q$ of 20. The measured laser-induced photocurrent amplitude (red dots) samples the spectral profile of the optical cavity

resonance. **e**, Electrical transfer (left) and output characteristics (right) of the device; the fit (red solid line) to the transfer data (open symbols) is explained in the main text.

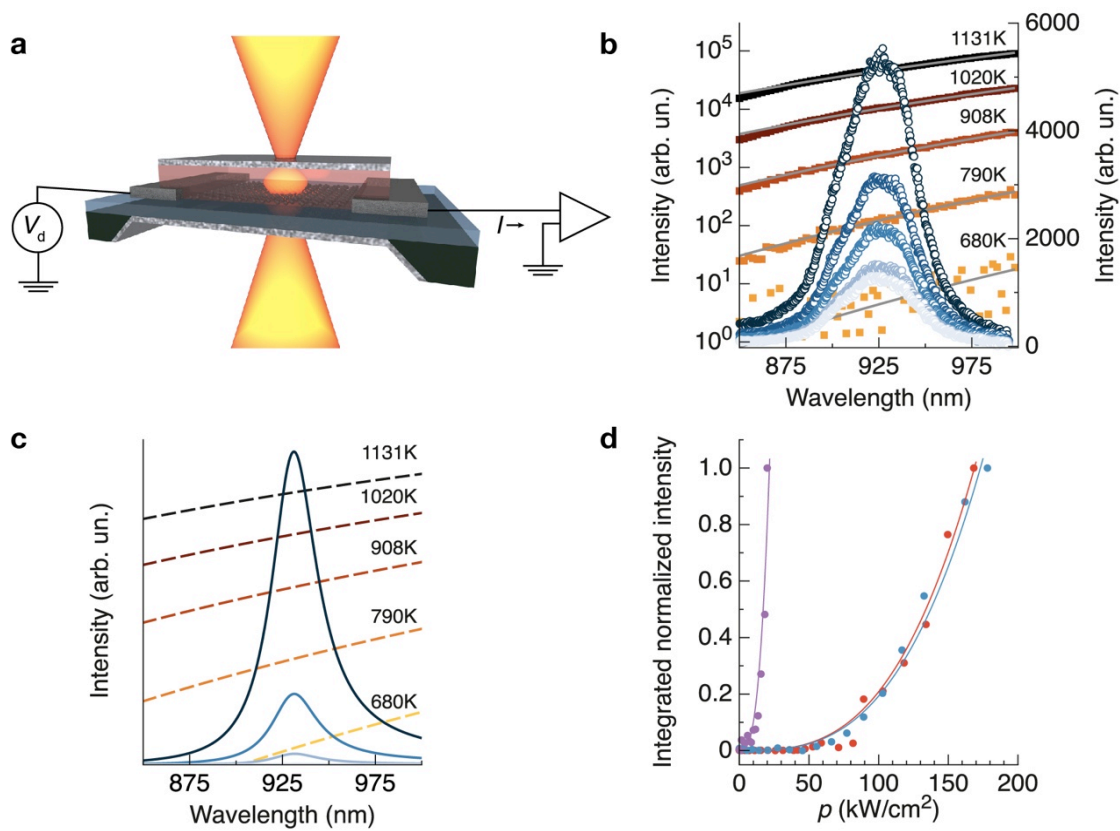

**Figure 3 | Electrically excited microcavity-controlled graphene thermal emitter.**
**a,** The schematic visualizes how thermal light emission is generated by applying a drain bias and how thermal radiation couples to the optical cavity mode. **b,** Thermal near-infrared emission spectra measured for a cavity-confined (open circles) and a non-confined (filled squares) graphene transistor. The emission spectra of the cavity-controlled graphene transistor displays the optical resonance of the cavity at $\lambda_{cavity}$=925nm. The left y-axis is the intensity of the non-confined emission (log scale) while the right y-axis is the intensity of the confined emission (linear scale). The indicated temperatures are derived by fitting the non-confined thermal radiation spectra to Planck's law (see Methods). **c,** Simulated spectra of cavity-controlled thermal radiation (solid lines) and non-confined thermal radiation (dashed lines) modeled by assuming that the cavity resonance is spectrally located at $\lambda_{cavity}$=925nm and has a spectral full width at half maximum of 30nm. **d,** Spectrally integrated light intensity as function of electrical power for three devices with different channel sizes (red 1x1μm$^2$, blue 2x2μm$^2$, purple 4x4μm$^2$). The solid lines are $T^3$-fits assuming that the dissipated electrical power is proportional to the temperature $T$ in the graphene sheet.

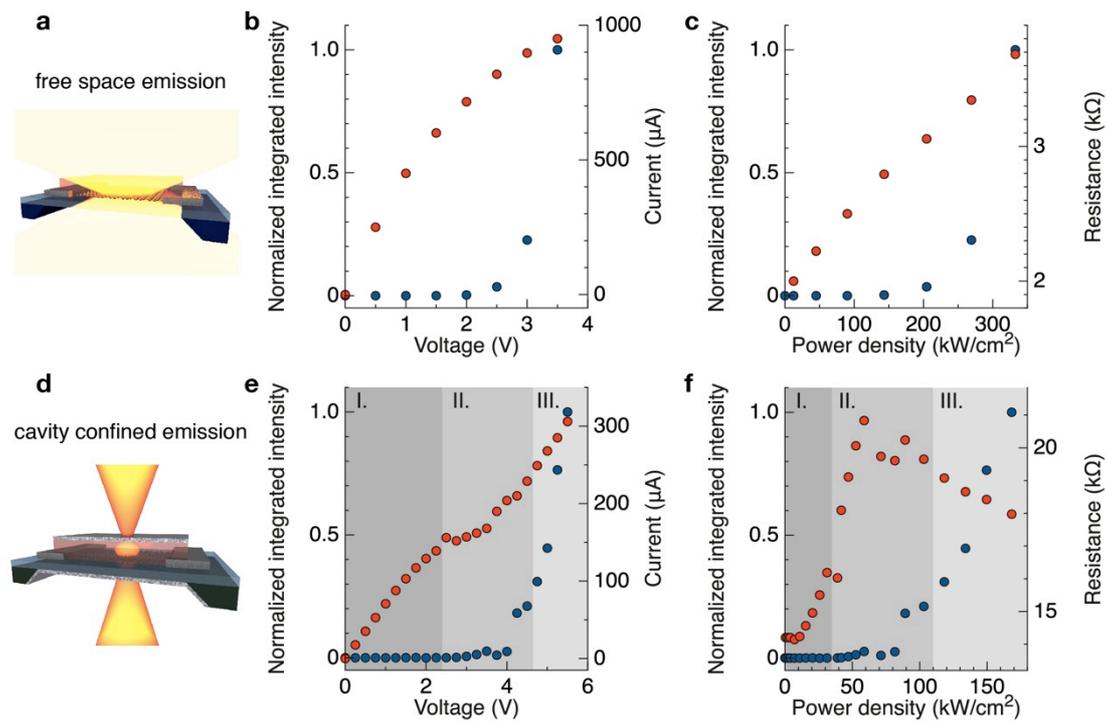

**Figure 4 | Optical confinement and electrical transport in a microcavity-controlled graphene transistor. a,** Schematic illustration of the sample layout. **b,** Normalized and integrated, emitted light intensity (blue) and electrical current (red) as a function of bias voltage measured with a non-confined graphene transistor and **c,** the integrated light intensity (blue) and the electrical resistance (red) plotted as a function of electrical power density. **d,** Schematic illustration of the sample layout. **e,** Normalized integrated emitted light intensity (blue) and electrical current (red) as a function of bias voltage measured with a microcavity-controlled graphene transistor and **f,** the integrated light intensity (blue) and the electrical resistance (red) plotted as

a function of electrical power density. Three regimes can be identified: (I.) sub-threshold, (II.) threshold and (III.) above-threshold, respectively.

**References**

1   Geim, A. K. and Novoselov, K. S., The rise of graphene. *Nat. Mater.* **6** (3), 183 (2007).

2   Geim, A. K., Graphene: Status and Prospects. *Science* **324** (5934), 1530 (2009).

3   Avouris, Ph., Graphene: Electronic and Photonic Properties and Devices. *Nano Lett.* **10** (11), 4285 (2010).

4   Bonaccorso, F., Sun, Z., Hasan, T., and Ferrari, A. C., Graphene photonics and optoelectronics. *Nat. Photon.* **4** (9), 611 (2010).

5   Xia, F. et al., Ultrafast graphene photodetector. *Nat. Nanotechnol.* **4** (12), 839 (2009).

6   Liu, M. et al., A graphene-based broadband optical modulator. *Nature* **474** (7349), 64.

7   Ju, L. et al., Graphene plasmonics for tunable terahertz metamaterials. *Nat. Nanotechnol.* **6** (10), 630 (2011).

8   Echtermeyer, T. J. et al., Strong plasmonic enhancement of photovoltage in graphene. *Nat. Commun.* **2**, 458 (2011).

9   Sun, Z. et al., Graphene Mode-Locked Ultrafast Laser. *ACS Nano* **4** (2), 803 (2010).

10  Nair, R. R. et al., Fine Structure Constant Defines Visual Transparency of Graphene. *Science* **320** (5881), 1308 (2008).

[11] Freitag, M. et al., Thermal infrared emission from biased graphene. *Nat. Nanotechnol.* **5** (7), 497 (2010).

[12] Berciaud, S. et al., Electron and Optical Phonon Temperatures in Electrically Biased Graphene. *Phys. Rev. Lett.* **104** (22), 227401 (2010).

[13] Fermi, E., Quantum theory of radiation. *Rev. Mod. Phys.* **4** (1), 87 (1932).

[14] Vahala, K. J., Optical microcavities. *Nature* **424** (6950), 839 (2003).

[15] Purcell, E. M., Spontaneous emission probabilities at radio frequencies. *Phys. Rev.* **69** (1-2), 37 (1946).

[16] Kleppner, D., Inhibited spontaneous emission. *Phys. Rev. Lett.* **47** (4), 233 (1981).

[17] Bjoerk, G. and Yamamoto, Y., in *Spontaneous Emission and Laser Oscillation in Microcavities*, edited by H. Yokoyama and K. Ujihara (CRC Press, Boca Raton, 1995).

[18] Steiner, M. et al., Controlling molecular broadband-emission by optical confinement. *New J. Phys.* **10** (12), 123017 (2008).

[19] Kim, S. et al., Realization of a high mobility dual-gated graphene field-effect transistor with Al2O3 dielectric. *Appl. Phys. Lett.* **94** (6), 062107 (2009).

[20] Lee, E. J. H. et al., Contact and edge effects in graphene devices. *Nat. Nanotechnol.* **3** (8), 486 (2008).

[21] Kim, H., Lim, S. C., and Lee, Y. H., Size effect of two-dimensional thermal radiation. *Phys. Lett. A* **375** (27), 2661 (2011).

[22] Perebeinos, V. and Avouris, Ph., Inelastic scattering and current saturation in graphene. *Phys. Rev. B* **81** (19), 195442 (2010).

[23] Novoselov, K. S. et al., Two-dimensional atomic crystals. *Proc. Natl. Acad. Sci.* **102** (30), 10451 (2005).

24    Casiraghi, C. et al., Rayleigh Imaging of Graphene and Graphene Layers. *Nano Lett.* **7** (9), 2711 (2007).

25    Ferrari, A. C. et al., Raman Spectrum of Graphene and Graphene Layers. *Phys. Rev. Lett.* **97** (18), 187401 (2006).

26    Wang, X., Tabakman, S. M. & Dai, H. Atomic Layer Deposition of Metal Oxides on Pristine and Functionalized Graphene. *Journal of the American Chemical Society* **130**, 8152–8153 (2008).

27    Hecht, E., *Optics*, 4. ed. (MA: Addison-Wesley Publishing Company, 2001).

**Supplementary Information**

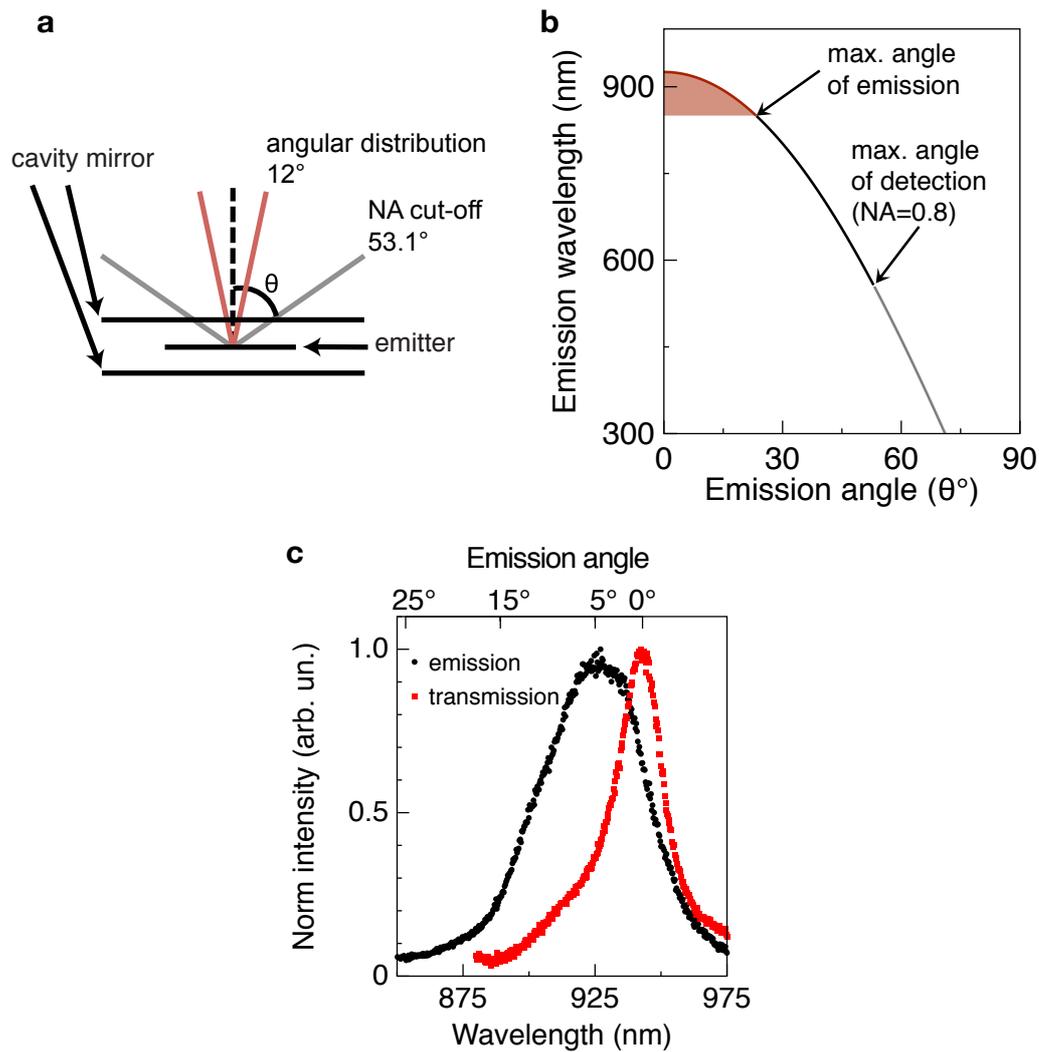

**Supplementary Figure S1 | Angular distribution of microcavity-controlled thermal emission. a,** Schematic showing the graphene layer placed between the two cavity mirrors. Also indicated are the maximum angle of detection and the angular distribution of the emitted light. **b,** Emission wavelength as a function of emission angle. No light emission can be observed for $\theta > 23°$. **c,** On-axis transmission (red) and thermal emission (black) spectrum measured on the same device as a function of wavelength and emission angle, respectively. The maximum thermal emission intensity is observed at 5° while the collimated thermal emission lobe has an angular width of 12° (FWHM).

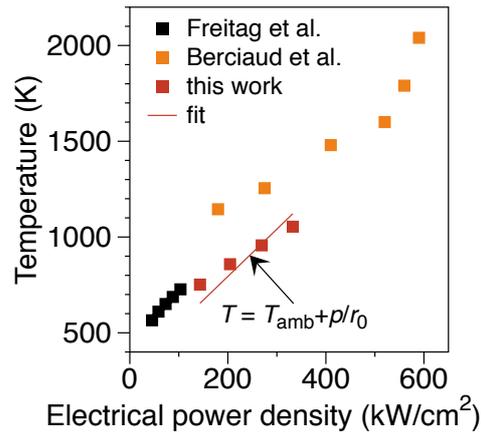

**Supplementary Figure S2 | Temperature dependence of non-confined, free-space graphene transistors.** Plotted is the temperature of graphene as function of injected electrical power density, in comparison with data taken from the literature[11,12]. The temperature-values are extracted by fitting the measured, free-space thermal emission spectra of a reference graphene transistor (with the same dielectric layers and metal contacts, but without cavity mirrors) to a model of a two-dimensional black-body.

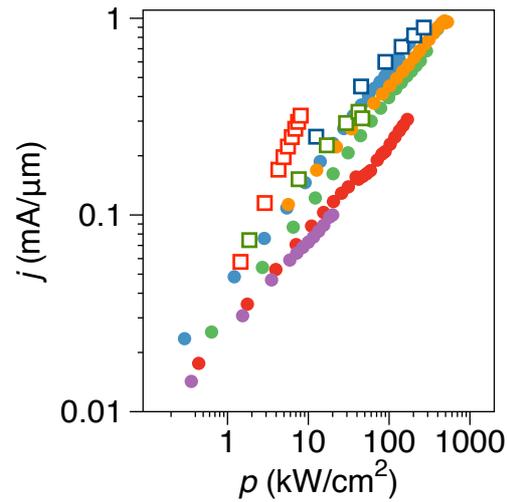

**Supplementary Figure S3 | Comparison of electrical current levels in confined and non-confined graphene transistors.** The experimental current density is plotted as function of electrical power density for a total of 8 graphene transistors. Microcavity-controlled graphene transistors (filled circles) consistently exhibit current saturation at lower electrical power densities than non-confined graphene transistors (open squares), along with a sudden increase of current density above threshold for thermal emission. Modifications of the electrical transport in microcavity-controlled graphene transistors that are in qualitative agreement with the data of the main text have been observed in a total of 5 devices.

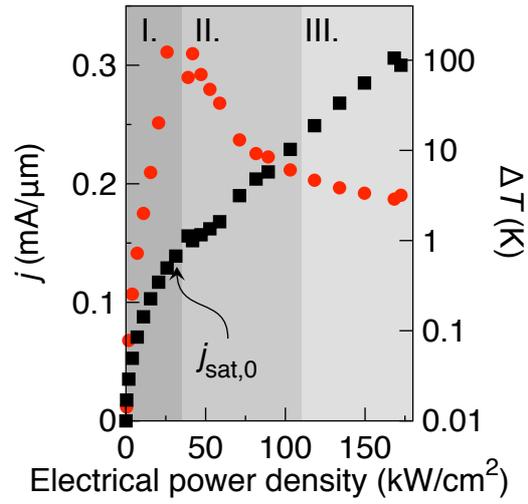

**Supplementary Figure S4 | Temperature modifications of biased, cavity-confined graphene.** Electrical current density (black squares) and relative temperature change (red circles) as function electrical power density. The cavity-confined graphene transistor heats up by $\Delta T$ as a consequence of the inhibition of thermal radiation (within regime I.). Once thermal radiation sets in (threshold regime II.), the temperature elevation decreases.

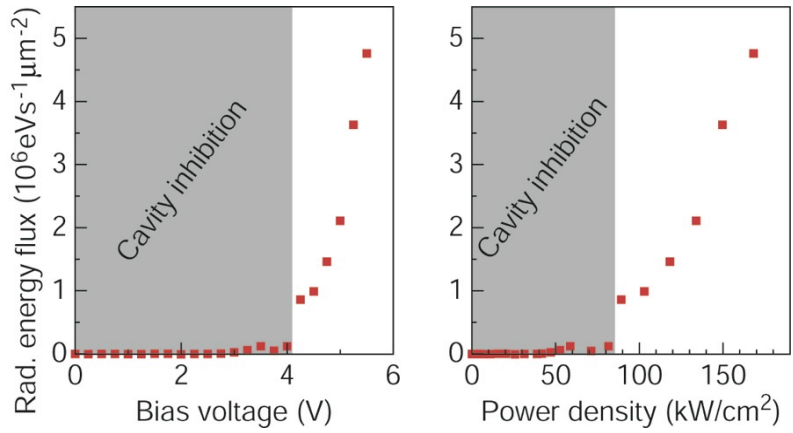

**Supplementary Figure S5 | Radiated heat flow as function of bias voltage and electrical power density.** Estimation at which rate heat is radiatively dissipated through cavity-controlled thermal light emission the measured light intensity is converted into an energy flux. By summing the measured photon flux in the spectral window of the cavity resonance and taking into account the detection conditions, we obtain the integrated energy flow that emanates from the device area of 1μm$^2$ at each bias point. Above threshold at 4V or 80kW/cm$^2$, respectively, the device emits thermal photons having energy of 1.3eV (λ~925nm) which is equivalent to an overall heat transfer at a rate of $4.8 \times 10^6$ eV/s.

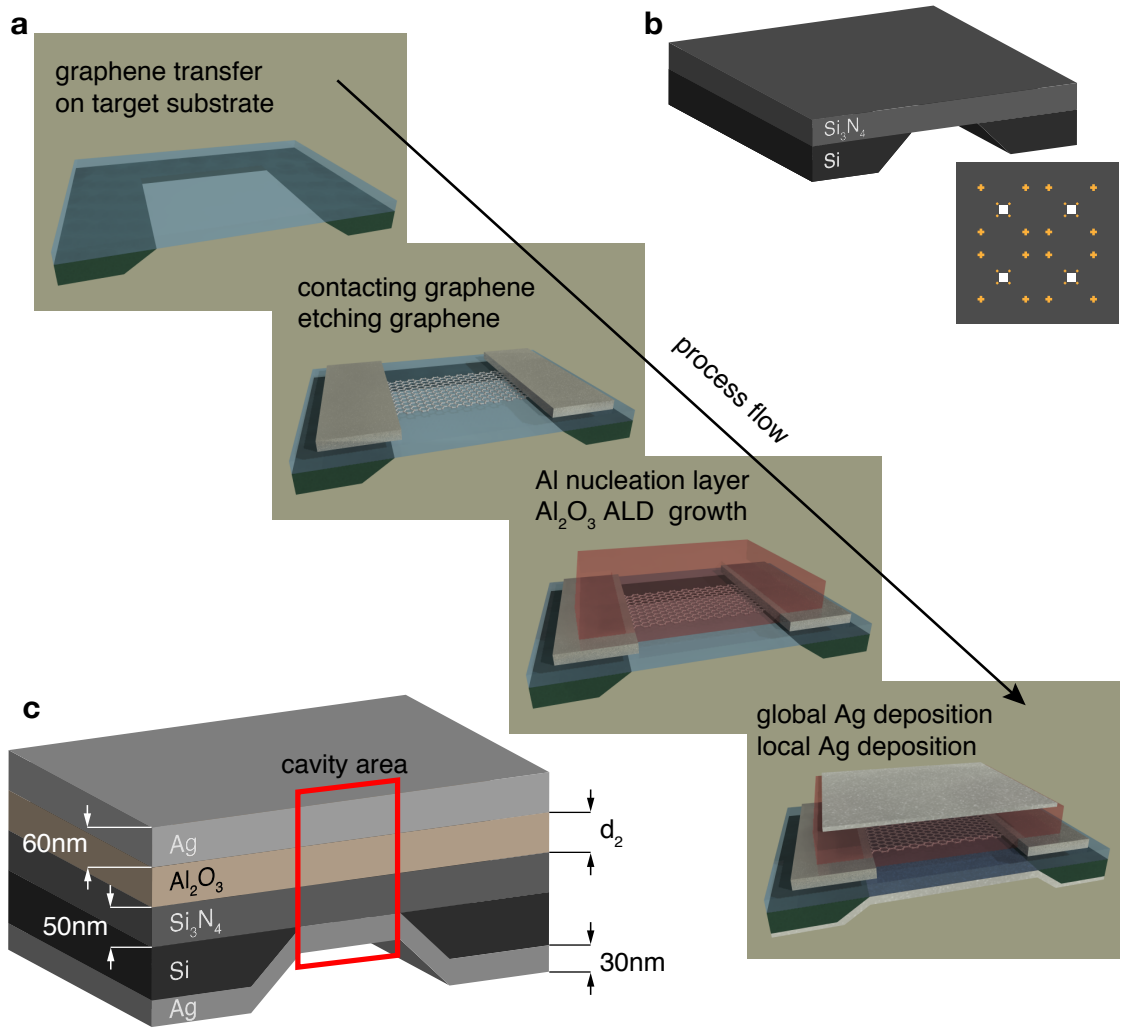

**Supplementary Figure S6 | Design and process flow for integrating a graphene transistors and an optical microcavity. a,** Process flow showing the main steps of monolithic integration. **b,** Schematic of the target substrate, a 50nm $Si_3N_4$ membrane with an area of $50\times50\mu m^2$. Also shown is a top view schematic with the predefined metallic markers defined by e-beam lithography used for orientation during the transfer process and re-alignment for later lithography steps. **c,** Schematic of the multilayer stack with detailing the different materials and layer thickness. The red box highlights the microcavity.

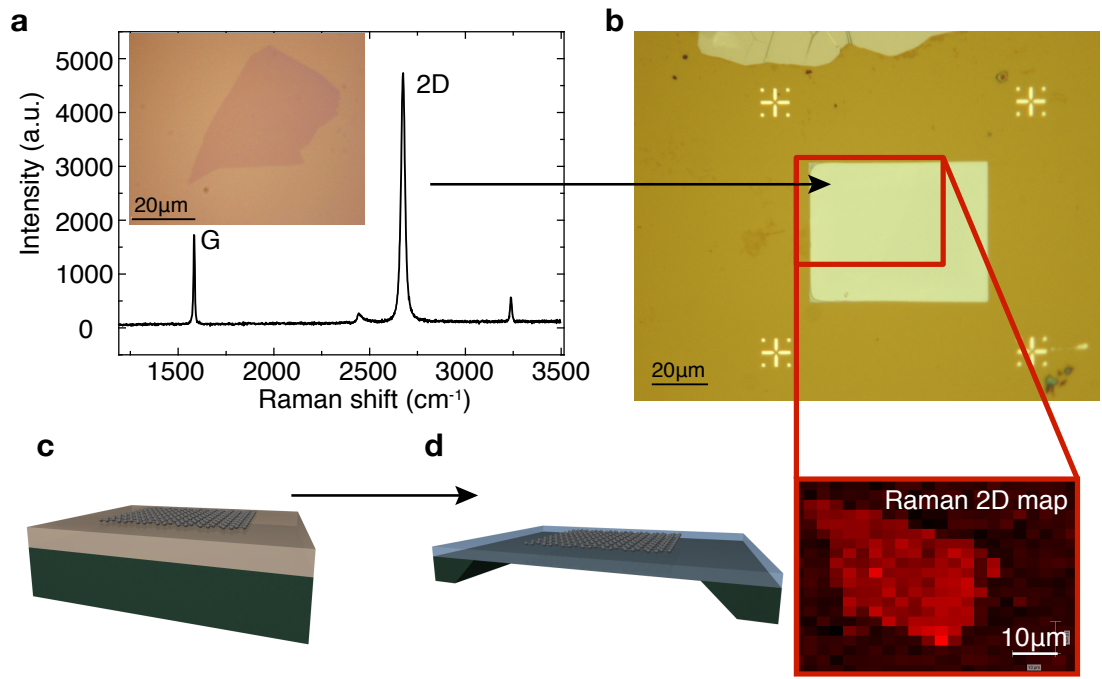

**Supplementary Figure S7 | Exfoliation and transfer of single layer graphene. a,** Raman spectrum of an exfoliated, single-layer graphene flake. Inset: Optical microscope image of the exfoliated single layer graphene on Si/SiO$_2$. **b**, Optical microscope image of single-layer graphene after transfer on to 50nm Si$_3$N$_4$ membrane. Also shown is a spatial map of the 2D Raman intensity confirming the successful transfer. **c** and **d,** 3D visualizations of single layer graphene after exfoliation and transfer on to the target substrate.

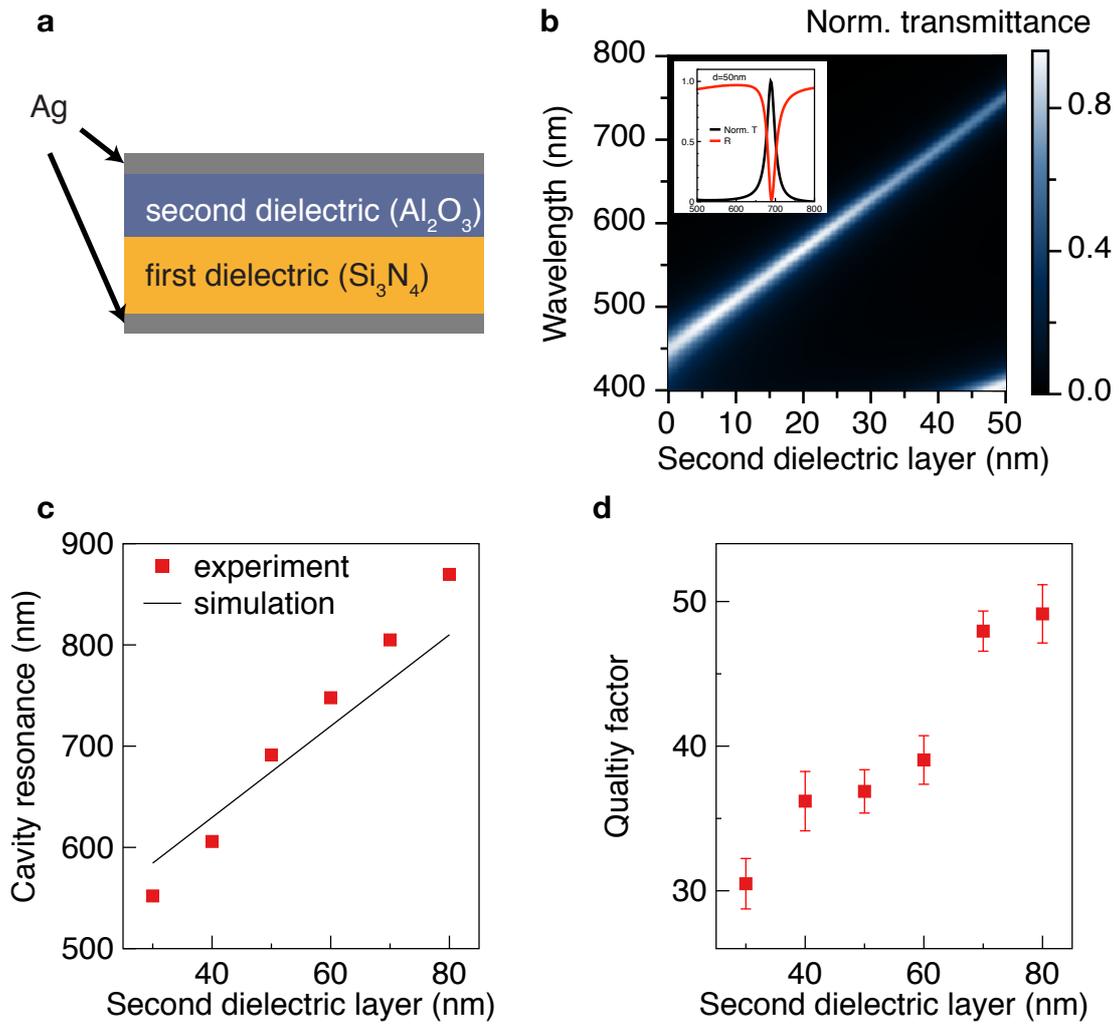

**Supplementary Figure S8 | Simulation and design of the optical microcavity. a,** Schematic of the multilayer stack used for modeling the transmission/reflectance of the microcavity as a function of wavelength and dielectric layer thickness. **b,** False color plot of the normalized transmittance from the cavity shown in **a**. **c,** Cavity resonance wavelength obtained by measuring the reflectance from reference cavities with $Al_2O_3$ layer thickness between 30nm and 80nm. Each data point represents the average of 4 different cavities. The solid line is a result from the optical simulation shown in **b**. **d,** Cavity-quality factors of the reference samples shown in c. Each data point represents the average of 4 different cavities.

**Supplementary Methods**

**Simulation and experimental verification of resonance wavelength and cavity-Q**

The microcavity resonance is determined by the thickness of the $Si_3N_4$ and $Al_2O_3$ layers, i.e. the total thickness of the intra-cavity medium. In our case the $Si_3N_4$ layer thickness is fixed at 50nm while the thickness of the second intra-cavity medium $Al_2O_3$ can be freely adjusted. To determine the $Al_2O_3$ thickness to grow, we modeled the cavity resonance within a transfer matrix approach, i.e. a plane wave propagates through this multilayered cavity system and the electric field is evaluated at each boundary (see Supplementary Figure S7a). From the simulations we infer the proper $Al_2O_3$ thickness for a targeted cavity resonance wavelength (Supplementary Figure S7b). For validation of the simulations we build a series of reference cavities without graphene. The results in Supplementary Figure S7c demonstrate the level of control over the cavity resonance by adjusting the thickness of the alumina layer. The cavity-$Q$ values achieved (see Supplementary Figure S7d) are in agreement with previous reports on planar cavities with metallic mirrors[28].

**Microcavity-controlled thermal emission – angular distribution**

We now analyze the angular distribution of the cavity-confined thermal light emission. According to reference 18, the following equation relates the emission wavelength $\lambda$ to the emission angle $\theta$ with respect to the cavity normal (see Supplementary Figure S1),

$$\lambda = \frac{2L n_{pol} \cos\theta}{(m - \Delta\phi/2\pi)}$$

where *L* is the geometrical mirror spacing (or cavity length), $n_{pol}$ is refractive index of the intra-cavity medium, *m* is the mode order (m=1 in this case), and $\Delta\phi$ is the phase shift associated with light absorption in the metallic mirrors. The microscope objective used in our measurements has a numerical aperture of NA=0.8. This corresponds to a maximum detection angle of $\theta_{max}$=53.1° with respect to the cavity normal (see Supplementary Figure S1). For the device discussed in the main manuscript (see Figs.3,4), having an on-axis ($\theta$=0) resonance wavelength of 925nm, the shortest wavelength that can be collected is hence 556nm. The measured microcavity-controlled emission spectra typically level of at around 850nm and have a spectral width of 40nm (FWHM), which translates into an angular distribution of $\theta$=12° (FWHM), with an intensity maximum occurring at $\theta_{max}$=5° (see Supplementary Figure S1). This demonstrates that the cavity-coupled thermal emission of graphene is radiated into a narrow lobe and that off-axis emissions coupled to guided modes are insignificant.

**Estimating temperature effects in a microcavity-controlled graphene transistor**

At high source-drain bias, the saturation current density $j_{sat}$ in graphene depends on the self-heating of the graphene layer[22]. The degree of self-heating is determined by the thermal coupling of graphene to its environment. A measure for the thermal coupling is the thermal conductance *r*. Within the self-heating model the saturation current is proportional to the square root of the thermal conductance,

$$j_{sat} \propto r^{0.5}$$

We extract a lower bound for the thermal conductance $r_0$ in our graphene transistors by fitting in Supplementary Figure S2 the measured spectra of the free space, non-confined thermal radiation to the following expression

$$T = T_{amb} + j \cdot F/r_0$$

which delivers $r_0$=0.4 kW/(cm$^2$K).

We now estimate the temperature modifications $\Delta T$ associated with the optical confinement based on the expression

$$\Delta T = \frac{j_{sat} F}{r}$$

We assume that $j_{sat}$=$j$-$j_{sat,0}$ for $j > j_{sat,0}$ and that relative changes in the thermal conductance $r$ can be captured through the relative changes of the saturation current

$$r = r_0 \left(\frac{j_{sat}}{j_{sat,o}}\right)^2$$

As reference saturation current $j_{sat,0}$, we choose the current density at the intersection between regimes I. and II. (see Supplementary Figure S4).

The resulting temperature modifications for the cavity emitter discussed in Figs. 3, 4 of the main manuscript are plotted in Supplementary Figure S4. As compared to graphene in free, non-confined space, the modifications of saturation current suggest temperature modifications as high as $\Delta T$=100K.

**Supplementary References**

28  Becker, H., Burns, S., Tessler, N. & Friend, R. Role of optical properties of metallic mirrors in microcavity structures. *J Appl Phys* **81**, 2825–2829 (1997).